\documentstyle[preprint,times,aps,prc,epsf,floats,tighten]{revtex}
\begin{document}
\preprint{SSF94-04-03 \hspace{0.5cm} to be published in Phys.Rev.C}
\title{Two-body currents in inclusive electron scattering}
\author{V. Van der Sluys, J.Ryckebusch\thanks{Postdoctoral research
fellow NFWO} \   and M.Waroquier\thanks{Research Director NFWO}\\
{\em  Laboratory for Nuclear
Physics}\\
{\em  Laboratory for Theoretical
Physics}\\
{\em Proeftuinstraat 86}\\
{\em B-9000 Gent, Belgium}}

\date{\today}

\maketitle

\begin{abstract}
\noindent The longitudinal and transverse structure functions
are calculated for
inclusive electron scattering from ${}^{12}$C and ${}^{40}$Ca
in the quasielastic and dip
region. The microscopic model presented here incorporates two-body
currents derived from one-pion exchange and intermediate $\Delta$
excitation. The reaction mechanism involves both one-nucleon and
two-nucleon knockout processes. It is demonstrated that, even for
quasielastic kinematics, two-body currents give a substantial
contribution to the transverse structure function.
Furthermore, the observed excess of transverse strength in the
 dip region between
 the quasielastic and delta peak can be partially ascribed to
direct two-nucleon
knockout following photoabsorption on a
two-body current.
\end{abstract}


\newpage

\section{Introduction}

During the last decade much experimental effort has been put into the
separation of the longitudinal ($R_L$) and transverse ($R_T$) structure
functions for
inclusive (e,e$'$) scattering from a number of nuclei
\cite{mez}\cite{bar}\cite{yat}\cite{dea}\cite{zgh}.
Whereas the quasifree (e,e$'$) cross section
could be
reasonably well described in the impulse approximation (IA)  within a
simple Fermi Gas Model (FGM)\cite{mon}, an accurate and simultaneous
description of the total cross section {\em and} separated structure functions
needs more
sophisticated theoretical models.

In the {\em quasielastic} (QE) {\em peak} , the transverse to longitudinal
ratio
of the response functions and  the quenching of
the longitudinal response function cannot be simultaneously reproduced within
the FGM.
Whereas the $R_T$ data are reasonably well described,
the longitudinal structure function is
systematically overestimated. Over the years, different
modifications to the FGM were suggested.
The longitudinal response function was shown to be sensitive to various
nuclear properties as there are: medium modified nucleon properties
\cite{nob}, final-state interactions (FSI),
random phase
approximation (RPA) correlations \cite{cav}\cite{alb} and relativistic effects
\cite{chin}\cite{hor}.
All these mechanisms are found to reduce the longitudinal strength thereby
improving  the agreement with the data.
Most of the aforementioned  corrections, however, affect the  transverse
structure function in the same way, worsening the agreement with the data.
As several many-body effects have been demonstrated to modify the
absolute (e,e$'$) response functions in the QE region and the different
approaches do not agree on their relative importance, an accurate description
of the $R_T/R_L$ ratio represents one of the major challenges  to
any theoretical approach. Another motivation for studying the  $R_T/R_L$
ratio in more detail is provided by the  findings of  a y-scaling
analysis of quasielastic scattering data \cite{finn}.
The predicted scaling behaviour of the longitudinal and transverse strength is
violated throughout the QE region pointing towards other reaction mechanisms
which are not incorporated in the adopted
nucleon-only approach.
Most studies dealing with the effect of many-body properties
 have one common feature~: they start from
the picture that the virtual photon couples with the individual nucleons
in the nucleus. As such, the
nuclear current is taken to be a one-body operator (impulse
approximation).
However, it has been demonstrated
\cite{car}\cite{koh}\cite{blu}
\cite{alb2}\cite{dek}\cite{ama}  that even in the
QE regime two-body mesonic currents (MEC) can induce considerable
corrections to the transverse structure function. In this context, Blunden {\em
et al}.
\cite{blu}
pointed out that in the QE region the one-nucleon knockout contribution
related to MEC
should be estimated as about
$10\%$ of the total strength.
Similar conclusions were drawn by J.Carlson and R.Schiavilla \cite{car}
for light nuclei.
The results of their microscopic calculation demonstrated
the essential role of virtual pion
exchange in the description of the quasielastic ${}^{4}$He(e,e$'$) data.

In the {\em dip region} between the QE peak and the $\Delta$ peak
an excess of transverse strength is observed experimentally \cite{bar}. For
moderate
values of the momentum transfer, pion  electroproduction is
estimated to be small in the high energy tail of the QE peak and is
unlikely to account for the missing strength \cite{bar}.
Over the years, the observed strength has been mainly attributed to two-nucleon
emission processes.
A  number of calculations have accounted for these two-nucleon knockout
processes
incorporating two-body currents within the FGM \cite{bar}\cite{dek}
\cite{ord}.
 It has been demonstrated
that part of the strength in the dip region originates from
two-particle knockout processes after photoabsorption on these two-body
currents.

In this paper we focus on inclusive electron scattering from medium-light
nuclei at the quasielastic peak and in the dip region
and report on a fully microscopic non-relativistic calculation based on
the continuum RPA model.
Due to the numerical complexity of these calculations, some restrictions have
been respected in the model. Most of them seem to be justified within the aim
of the present paper.
We summarize  the different ingredients and limitations of our model:
\begin{itemize}
\item{In the energy region under consideration,
the inclusive (e,e$'$) strength is assumed to originate solely from
one and two-body knockout processes.
As such, the nuclear charge-current four vector
is assumed to be the sum of a one and two-body part
related to one-pion exchange and
intermediate $\Delta$ excitation.
Real pion electroproduction is not incorporated in our approach. }
\item{At the quasielastic peak, our model goes beyond the direct nucleon
knockout approach and incorporates
RPA type of nucleon correlations within a
continuum HF-RPA formalism. Damping effects due to higher order excitations
of the $n$p-$n$h type ($n \geq 2$) are taken into account
 in a phenomenological way by introducing a complex self-energy.}
\item{No short-range correlation (SRC) corrections to the wave functions
 are implemented
in the model.}
\item{The model does not include any relativistic correction. Negative-energy
contributions tend to suppress both the longitudinal and the transverse
structure function to the same degree \cite{chin}.}
\end{itemize}
This paper is built up as follows.
The different aspects of the theoretical model are outlined
in section II. The results for inclusive electron scattering from $^{12}$C and
$^{40}$Ca are presented in section III. Some conclusions are drawn
in section IV.

\section{The model}

In the one-photon exchange approximation the (e,e$'$) cross section reads as
\begin{eqnarray}
\label{cross}
\frac{{\rm d}^{3}\sigma}{{\rm d}E_{f}{\rm d}\Omega_{E_{f}}}(e,e')  &=&
\sigma_{M} \left\{\left(\frac{q_{\mu}^{2}}{\vec{q}^{2}}\right)^2
R_{L}(\vec{q},\omega) +
\left(\tan^2(\theta_e/2)-\frac{q_{\mu}^2}{2\vec{q}^{2}}\right)
R_{T}(\vec{q},\omega)
\right\} \;\; ,
\end{eqnarray}
where $q_{\mu} (\omega,\vec{q})$ is the four momentum transferred to the
nucleus and  $\sigma_{M}$ is the Mott cross section for scattering from a
point particle: $\sigma_{M}=\frac{\alpha^2 {\rm cos}^2(\theta_e/2)}
{4 E_i^2 {\rm sin}^4(\theta_e/2)}$. All information concerning the
electromagnetic structure
of the nucleus that can be derived from (e,e$'$) reactions,
 is contained in the longitudinal $R_{L}$ and transverse
$R_T$ structure functions. Both structure functions are related to the
matrixelements of the nuclear electromagnetic charge-current operator
$(\rho,\vec{J})$ in the following way
\begin{eqnarray}
\label{rlrt}
R_{L}(\vec{q},\omega)&=&\sum_{n} \mid <n\mid \rho(\vec{q}) \mid
0> \mid ^2 \delta (\omega-E_n+E_0)\;\; ,\qquad\qquad\qquad \\
R_{T}(\vec{q},\omega)&=&\sum_{n} \left\{ \mid <n\mid J_{+1}(\vec{q}) \mid
0>\mid ^2 + \mid <n\mid J_{-1}(\vec{q}) \mid
0> \mid ^2 \right\} \delta (\omega-E_n+E_0)\;\;.
\end{eqnarray}
The sum over $n$ extends over all final nuclear states $\mid n >$ with
excitation energies $(E_n-E_0)$ relative to the groundstate energy $E_0$ of
the target nucleus $\mid 0 >$ ($J^{\pi}=0^+$).
In our approach, the sum over all final states  includes one and two-body
knockout processes.
To be more specific, we  assume the inclusive (e,e$'$)
 strength to be mainly originating  from (e,e$'$N) and (e,e$'$2N) processes
i.e.
\begin{eqnarray}
\label{cross2}
\frac{{\rm d}^{3}\sigma (e,e')}{{\rm d}E_{f}{\rm d}\Omega_{E_{f}}}  &=&
\sum_{N}\int {\rm d}\Omega_N  {\rm d}T_N
\frac{{\rm d}^{6}\sigma (e,e'N)}{{\rm d}E_{f}{\rm d}\Omega_{E_{f}} {\rm d}
\Omega_N {\rm d}T_N}\nonumber\\
&+&\sum_{N,N'} \int {\rm d}\Omega_N  {\rm d}\Omega_{N'} {\rm d}T_N
 {\rm d}T_{N'}\frac{{\rm d}^{9}\sigma (e,e'N N')}
{{\rm d}E_{f}{\rm d}\Omega_{E_{f}} {\rm d}
\Omega_N {\rm d}T_N {\rm d}
\Omega_{N'} {\rm d}T_{N'}}\;\; .
\end{eqnarray}
In this expression $N$ and $N'$ stand for all occupied proton and neutron
single-particle (s.p.)
states of the A-particle nucleus. In order to estimate the contribution
from these processes
to the inclusive cross sections, the above expression involves an
integration  over the solid
angles and energies of the outgoing particles. The kinetic energy
 of the outgoing nucleon(s) is fixed by the energy conservation relation
$\omega=
E_x^{A-1}+S_{N}+T_N+T_{A-1}$ for one-body knockout and $\omega=
E_x^{A-2}+S_{2N}+T_N+T_{N'}+T_{A-2}$ for two-body knockout. The
excitation energy of the residual nucleus and the one-nucleon separation
energy are denoted by $E_x$ and $S_N$.
In line with our model assumptions, the structure functions consist of a one-
and two-nucleon
knockout part, i.e.
$R(\vec{q},\omega)\equiv R^{[1]}(\vec{q},\omega)+R^{[2]}(\vec{q},\omega)$.

\begin{figure}
\hspace{2.2cm}
\epsfxsize=12cm
\epsffile{diag.eps}
\caption{Diagrams considered in the (e,e$'$N) cross section:
(a) impulse approximation, (b) one-pion exchange contribution.
Two-body knockout following photoabsorption on a two-body current is
depicted by diagrams of the type (c).
For the two-body current contributions only pion-in flight
 diagrams are displayed. }
\end{figure}

The
wave function for one escaping particle and a residual $(A-1)$-nucleus
(Fig.~1a) is evaluated in the continuum RPA formalism as
described in ref.\cite{jan}. The RPA formalism involves a partial-wave
expansion
of the final state in terms of linear combinations of
particle-hole and hole-particle excitations out of a correlated A particle
 groundstate.
Bound and continuum single-particle states are  eigenstates
of the Hartree-Fock  (HF) mean-field potential obtained with an effective
interaction of
the Skyrme type (SKE2)\cite{war}. In this way, the bound and the continuum
state wave functions
remain orthogonal. In terms of the FSI a continuum RPA calculation is
equivalent with a coupled-channel calculation in which one-proton and
one-neutron emission from the different shells is implemented.

By analogy with the one-nucleon emission picture, the wave function for
two escaping particles and a (A-2)-residual nucleus is obtained by
performing a double partial-wave expansion in terms of 2p-2h states
\cite{jan3}.
The wave functions for the two-particle (2p) continuum states are
evaluated in the same HF mean field potential as for the one-body emission
case.
In this way, we arrive at a consistent description of the one- and
two-nucleon knockout cross sections. It should be stressed, however,
 that in the
two-nucleon knockout calculation a direct knockout reaction mechanism is
assumed and no channel couplings are implemented.

In the present approach, the (e,e$'$N) and (e,e$'$2N)
reaction mechanism  encompasses one-nucleon and two-nucleon knockout
   following photoabsorption on a one-body and two-body current.
As such, the nuclear current consists of a
one-body part determined by the convection and
magnetization current and a two-body part related to one-pion exchange.
The two-body current is taken from a non-relativistic reduction of the
lowest-order Feynman diagrams with one exchanged pion and intermediate
delta excitation. We adopt pseudovector coupling of the pion to the
nucleon. This procedure gives rise to the seagull terms, the
pion-in-flight term and terms with a $\Delta(1232)$ excitation in the
intermediate state. To lowest order, the nuclear charge
operator is not affected by two-body contributions.
Consequently, within our model assumptions all longitudinal strength originates
from
photoabsorption on a one-body operator. The adopted two-body
charge-current four vector is taken from ref. \cite{riska} and reads :
\begin{eqnarray}
\label{current}
\rho^{(2)}(\vec{q};\vec{q}_{1}\vec{q}_{2})&=&0 \nonumber\\
\vec{J}^{(2)}(\vec{q};\vec{q}_{1}\vec{q}_{2})&=&
\vec{J}^{(2)}_{seag}(\vec{q};\vec{q}_{1}\vec{q}_{2})
+\vec{J}^{(2)}_{pion}
(\vec{q};\vec{q}_{1}\vec{q}_{2})+
\vec{J}^{(2)}_{delta}(\vec{q};\vec{q}_{1}\vec{q}_{2}),
\end{eqnarray}
with
\begin{eqnarray}
\label{seag}
\vec{J}^{(2)}_{seag}(\vec{q};\vec{q}_{1}\vec{q}_{2})&=&
-ie\left(\frac{f_{\pi NN}}{m_{\pi}}\right)^{2}
F_{\gamma N}(q)(\vec{\tau}_{1}\times\vec{\tau}_{2})^{3}
\left\{ \frac{\vec{\sigma}_{1}(\vec{\sigma}_{2}.\vec{q}_{2})}
{\vec{q}_{2}^{2}+m_{\pi}^{2}}-
\frac{\vec{\sigma}_{2}(\vec{\sigma}_{1}.\vec{q}_{1})}
{\vec{q}_{1}^{2}+m_{\pi}^{2}}\right\}, \nonumber \\
\vec{J}^{(2)}_{pion}(\vec{q};\vec{q}_{1}\vec{q}_{2})&=&
ie\left(\frac{f_{\pi NN}}{m_{\pi}}\right)^{2}
F_{\gamma\pi}(q)(\vec{\tau}_{1}\times\vec{\tau}_{2})^{3}
\frac{(\vec{\sigma}_{1}.\vec{q}_{1})(\vec{\sigma}_{2}.\vec{q}_{2})}
{(\vec{q}_{1}^{2}+m_{\pi}^{2})(\vec{q}_{2}^{2}+m_{\pi}^{2})}
(\vec{q}_{1}-\vec{q}_{2}),\nonumber \\
\vspace{.3cm}
\vec{J}^{(2)}_{delta}(\vec{q};\vec{q}_{1}\vec{q}_{2})&=&i
\frac{f_{\gamma
 N\Delta}f_{\pi N\Delta}f_{\pi NN}}{9m_{\pi}^{3}}
\left(\frac{1}{M_{\Delta}-M_{N}-\omega-\frac{i}{2}
\Gamma_{\Delta}(\omega)}+\frac{1}{M_{\Delta}-M_{N}+\omega}\right)
F_{\gamma\Delta}(q) \nonumber\\ &&
 \left\{ 4\frac{(\vec{\sigma}_{2}.\vec{q}_{2})}
{\vec{q}_{2}^{2}+m_{\pi}^{2}}\vec{q}_{2}\tau_{2}^{3}
 -(\vec{\tau}_{1}\times\vec{\tau}_{2})^{3}\left(
\frac{\vec{\sigma}_{2}.\vec{q}_{2}}
{\vec{q}_{2}^{2}+m_{\pi}^{2}}(\vec{\sigma}_{1}\times
\vec{q}_{2})\right) + [1\longleftrightarrow 2] \right\}\times
 \vec{q}.\nonumber
\end{eqnarray}
The following coupling constants are adopted
$\frac{f_{\pi NN}^{2}}{4\pi}=0.079$, $\frac{f_{\pi
N\Delta}^{2}}{4\pi}=0.37$ and $f_{\gamma N\Delta}^{2}=0.014$.
In the evaluation of the $\Delta$-current an
energy-dependent decay width $\Gamma_{\Delta}$
\begin{eqnarray}
\Gamma_{\Delta}(\omega)=\frac{8 f_{\pi NN}^{2}}{12 \pi}
  \frac{(\omega^2-m^2_{\pi})^{3/2}}{m^2_{\pi}}
\frac{M_{\Delta}-M_N}{\omega} \;\;
\end{eqnarray}
has been introduced \cite{ose}.

To account for the composite structure of the different vertices, the two-body
nuclear current is modified
by electromagnetic and hadronic formfactors. For the $\gamma N$ formfactor
($F_{\gamma N}$)  we use the
common dipole form \cite{gal}. The pion formfactor $F_{\gamma\pi}$ is
extracted from the vector dominance model \cite{eric}.
It should be stressed that the use of two different parameterizations
for the pion and nucleon formfactor violates current conservation.
An alternative choice that preserves current conservation would be
replacing the pion formfactor with the nucleon formfactor.
 Since for the energy-momentum region considered here both
formfactors differ at most $20\%$, either of both choices will not
appreciably affect the results.
A similar conclusion was drawn by Amaro {\em et al.} \cite{ama} in a
recent paper on the role of meson-exchange currents
in quasielastic electron scattering
from complex nuclei.
In addition, in ref. \cite{vsl94} we have
shown that the calculated contribution from MEC
to the quasielastic (e,e$'$p) structure functions
is  rather insensitive to the choice of the pion formfactor.
The delta current is divergenceless
and can be multiplied with an arbitrary formfactor $F_{\gamma\Delta}$.
 As is commonly done, we assume
$F_{\gamma\Delta}=F_{\gamma N}$ in all calculations presented here. The
short-range corrections to the $\pi NN$ and $\pi N
\Delta$  vertices
are implemented in a phenomenological way by introducing hadronic
formfactors. These formfactors are usually parameterized in momentum
space as
$(\Lambda_{\pi}^2-m^2_{\pi})/(\Lambda_{\pi}^2+\vec{p}^2)$ with
$\Lambda_{\pi}$ a scale parameter for the high-momentum cut-off.
Standard values of $\Lambda_{\pi}$ lie in the range $800-1250$ MeV.
All results presented in this paper are derived with $\Lambda_{\pi}$ set
equal to $1200$ MeV. This value is suggested by recent parameterizations of
the Bonn potential \cite{mach}.

The one-nucleon knockout channels can be fed by both the one- and
 two-body part of the nuclear current. In Fig.~1 a,b we depict the
diagrams that are included  in our model for one-nucleon emission.
The nuclear charge-current is expanded in terms of the Coulomb
$M^{coul}_{JM}(q)$, electric $T^{el}_{JM}(q)$ and magnetic
 $T^{mag}_{JM}(q)$
transition operators. In combination with the adopted
partial-wave expansion for the one-body knockout wave function,
 the calculation of the (e,e$'$N) cross section is
then reduced to evaluating matrixelements of the type :
\begin{eqnarray}
&&<(p(\epsilon_p lj)h^{-1});J\omega\mid\mid M^{coul (1)}_{J}(q)\mid\mid
0^{+}> \;\; , \label{trans1}
\\
&&<(p(\epsilon_p lj)h^{-1});J\omega\mid\mid T^{el,mag (1)}_{J}(q)+T^{el,mag
(2)}_{J}(q)\mid\mid
0^{+}> \;\; . \label{trans2}
\end{eqnarray}
 In this expression  $M^{coul (1)}_{J}$, $T^{el,mag (1)}_{J}$ refer to
 the one-body
and $T^{el,mag (2)}_{J}$ to the two-body part of
the transition operator.  The residual
nucleus is considered to remain in a pure hole state $h$  relative to
the groundstate of the target nucleus. The continuum state $p(\epsilon_p lj)$
satisfies  $\epsilon_{p}=\omega - \mid\epsilon_{h}\mid
> 0$, where $\epsilon_{h}$ is the hole single-particle energy as derived
from a HF calculation.
The two-body part of the transition operators is handled
without any further approximation
and  involves two active nucleons in the absorption process (Fig.~1b).
Hence, the evaluation of the two-body current part in the matrixelement
(\ref{trans2}) is reduced to
\begin{eqnarray}
\label{two}
&&<(p(\epsilon_p lj)h^{-1});J\omega\mid\mid T^{el,mag (2)}_{J}(q)\mid\mid
0^{+}>
 =  \sum_{h' J_{1}J_{2}} \sqrt{2J_{1}+1}\sqrt{2J_{2}+1}(-1)^{j_{h}-j_{h'}
-J-J_{2}}\nonumber\\
&&\qquad\qquad \times \left\{ \begin{array}{ccc}j_{h} &j_{h'}& J_{1}\\J_{2} &J&
j \end{array}
\right\}
<(h h');J_{1}\mid\mid T_{J}^{el,mag (2)}(q)\mid\mid (ph')
;J_{2}>_{as} \;\; ,
\end{eqnarray}
where the sum over $h'$ extends over all occupied s.p. states in the target
nucleus.
The antisymmetrized two-body matrix
 elements  with the pionic  currents can be found in ref. \cite{jan3}.

In the foregoing discussion  only diagrams of the type depicted in
Fig.~1 a,b are incorporated in the one-body emission model.
The coupling of 1p-1h excitations to 2p-2h and higher order excitations
is not considered.
However, it has been demonstrated in ref.\cite{co} that damping effects
resulting from these higher order contributions can be partially taken into
account in a
phenomenological approach by introducing a
complex self-energy for the p-h state
$\Sigma(\omega)=\Delta(\omega)+i\Gamma(\omega)/2$.
The one-body response is then derived using a folding procedure i.e.
\begin{eqnarray}
R^{[1]}(\vec{q},\omega)&\longrightarrow&\int_{0}^{\infty} dE
R^{[1]}(\vec{q},E)
\rho(E,\omega) \;\; .
\end{eqnarray}
The spreading function $\rho(E,\omega)$ is expressed as a function of
the real and imaginary part of the p-h self-energy:
\begin{eqnarray}
\rho(E,\omega)&=&\frac{(1/2\pi)\Gamma (\omega)}
{(E-\omega-\Delta(\omega))^{2}+(\Gamma(\omega)/2)^2} \;\;.
\end{eqnarray}
For  the p-h spreading width $\Gamma$ and energy shift $\Delta$,
we consider the same conventions and expressions as in ref. \cite{co}.

In contrast with the one-nucleon knockout process to which both one- and two-
body absorption mechanisms contribute,
two-nucleon emission can only
take place
 through photoabsorption on a two-body current within our model .
In Fig.~1c one of
the considered diagrams is depicted. The residual
(A-2)-nucleus is a pure 2h-state $\mid(hh')^{-1};J_RM_R>$ with respect to the
groundstate of the target nucleus. We have to evaluate matrixelements of
the two-body transition operators of the following type
\begin{eqnarray}
\label{trans3}
<\left[(hh')^{-1}J_R;(p(\epsilon_p lj),p'(\epsilon_{p'}l'j'))J_1\right];
J\omega\mid\mid
T_{J}^{el,mag (2)}(q)\mid\mid 0^{+}>.
\end{eqnarray}
The continuum particle states $p(\epsilon_p lj)$ and $p'(\epsilon_{p'}
l'j')$ satisfy $\epsilon_p+\epsilon_{p'} = \omega-\mid\epsilon_{h}\mid
-\mid\epsilon_{h'}\mid > 0$.

\section{Results for ${}^{12}$C(\protect{$e$,$e$}$'$) and
${}^{40}$Ca(\protect{$e$,$e$}$'$)}

In this paper we concentrate on inclusive electron
scattering from ${}^{12}$C and ${}^{40}$Ca. We
performed calculations for several values of the
momentum transfer i.e. ($^{12}$C(e,e$'$) :
 $q=400$ MeV$/$c , $q=550$ MeV$/$c) and
($^{40}$Ca(e,e$'$) : $q=370$ MeV$/$c).
The theoretical predictions for the two structure functions
of $^{12}$C(e,e$'$) are displayed in Figs.~2 and 3 and
are compared with the results of a Rosenbluth separation
 from Barreau {\em et al.} \cite{bar}. The calculations
for $^{40}$Ca(e,e$'$) are confronted with two different sets of data
\cite{mez}\cite{yat} and are depicted
in Fig.~4 .
In the analysis of the experiments outlined in refs.\cite{mez}, \cite{bar} and
\cite{yat},
all data have been corrected for Coulomb distortion effects adopting
the effective
momentum approximation (EMA).

In particular, for the longitudinal $^{40}$Ca(e,e$'$) structure function at
$q=370$ MeV$/$c we observe a severe mismatch between the MIT and Saclay
data. According to ref. \cite{yat} this inconsistency originates from an
error in the initial data taking with one of the two experimental set-ups and
the discrepancy does not lie in the adopted Rosenbluth separation procedure.
Further experimental investigation is highly needed to settle the
inconsistency between both data sets.

Firstly, in comparison with the Saclay data,
it turns
out that the qualitative behaviour  of the calculations is similar for
both  nuclei at different momentum transfers.
The results of the IA calculations are shown as
dashed lines in Figs.~2-4. These IA calculations encompass
one-nucleon emission after photoabsorption on a one-body current
and include the effect of 1p-1h RPA correlations and
additional spreading corrections.
At the quasielastic peak, the
Saclay longitudinal structure functions  for both target nuclei
are clearly overestimated by the
one-body knockout picture
whereas
for the transverse structure function the impulse approximation slightly
underestimates the experimental results.
Consequently, this is reflected in an inadequate  description of the
ratio $R_T/R_L$ as obtained from the Saclay data.
On the other hand, at the quasielastic peak, the MIT data for
the $^{40}$Ca(e,e$'$) reaction seem to be
reasonably reproduced in this nucleon-only picture and
any further extension
of the theoretical approach risks to worsen the agreement reached.
In the dip region, however,
the IA is inadequate to reproduce the
measured transverse strength for all data sets.

Inclusion of two-body currents in the model affects this picture
 in a drastic way.
{}From eq. (\ref{trans1}) it is clear that
the longitudinal structure function remains unaffected by the two-body
part in the nuclear current. On the other hand,
mesonic currents contribute substantially to the one-nucleon
knockout transverse cross section.
This enhancement
is relatively more important for the smallest momentum transfers
($q=400$ MeV$/$c for $^{12}$C, $q=370$ MeV$/$c for $^{40}$Ca)
considered here.
In comparison with the Saclay data, after including two-body currents the
calculations overestimate the measured longitudinal and
transverse strength to the same degree.  We want to stress here
that, given
the complexity of the calculations when including two-body nuclear
currents, we did not attempt to account for additional many-body effects,
like SRC \cite{trai2}\cite{tak} and relativistic effects \cite{chin}\cite{blu}.
All these corrections tend to
reduce both response functions to a more
or less similar degree, thus
leaving the transverse to longitudinal ratio almost unaffected.

The $^{40}$Ca(e,e$'$) data as measured in MIT-Bates exhibit a totally different
behaviour compared to the corresponding Saclay values. It is clear from Fig.~4
that at the quasielastic peak
the measured $R_T/R_L$ ratio is not
in favour of two-body currents.

In the past, the impact of two-body currents on the (e,e$'$) structure
functions was mostly investigated within the FGM. Only recently,
microscopic calculations that account for one-pion exchange currents have
become available.
We compared
 our results with the theoretical predictions by J.Carlson {\em et al}
\cite{car} for the $^{4}$He(e,e$'$) reaction
  and with a calculation similar in nature to ours
 of Amaro
{\em et al.}\cite{ama} ($^{12}$C(e,e$'$) and $^{40}$Ca(e,e$'$)).
For the latter the effect of
pion-exchange currents and the $\Delta$ current on the quasielastic transverse
structure function was investigated in a shell model framework which involves
1p-1h and 2p-2h nuclear final states.
 The impact of two-body currents on the transverse (e,e$'$) structure
function in the QE peak
 was found to be very different for
these two approaches.  In line with our results concerning the relative
importance of MEC,
virtual pion-exchange was established to be
a significant source of transverse $^{4}$He(e,e$'$) strength.
On the contrary, Amaro {\em et al.} found that the contribution of two-body
currents to the one-nucleon
knockout channel in $^{12}$C(e,e$'$) and $^{40}$Ca(e,e$'$) is negligible.
They attribute this small effect of two-body currents to the lack of SRC in
their nuclear wave functions. Notwithstanding the fact that
SRC are also neglected in our model,
 we find that a considerable amount of transverse
strength at the quasielastic peak can be ascribed to the MEC.
Explicit inclusion of the SRC effects is expected to have a reducing effect
on the strength produced by the meson-exchange currents. In ref. \cite{marc} it
was shown that this reduction is relatively small and of the order of $10\%$.

Two-particle
knockout comes into play beyond $\omega=100$ MeV.
{}From Figs.~2-4  it is clear that part of the transverse strength in
the dip region can be ascribed  to two-nucleon emission processes
following electro-induced
photoabsorption on the mesonic currents.
The q-dependence for the calculated 2N knockout strength
is similar to the one observed for
the mesonic contribution to the one-body
knockout strength. The two-nucleon knockout
strength decreases with increasing momentum transfer.
Despite the large
experimental error
bars for the transverse to longitudinal ratio in the dip region, our model
seems to describe reasonably well the $\omega$ dependence of $R_T/R_L$.
This is clear evidence for the importance of including two-particle knockout
when describing the inclusive cross section in the dip region.
In contrast with the results for the one-nucleon knockout channel,
our model predictions for the
two-nucleon knockout contribution to the transverse structure
functions are consistent
with those of Amaro {\em et al.} \cite{ama} .

It has to be emphasized that our main focus in this study was on
 the quasielastic
and dip region of the inclusive (e,e$'$) spectrum.
As mentioned before, real pion electroproduction is neglected in
our model. Consequently we fail in describing the high $\omega$ side of
the measured transverse strength which is expected to be mainly originating
from these processes \cite{dek}.

\begin{figure}
\hspace{3cm}
\epsfxsize=10cm
\epsffile{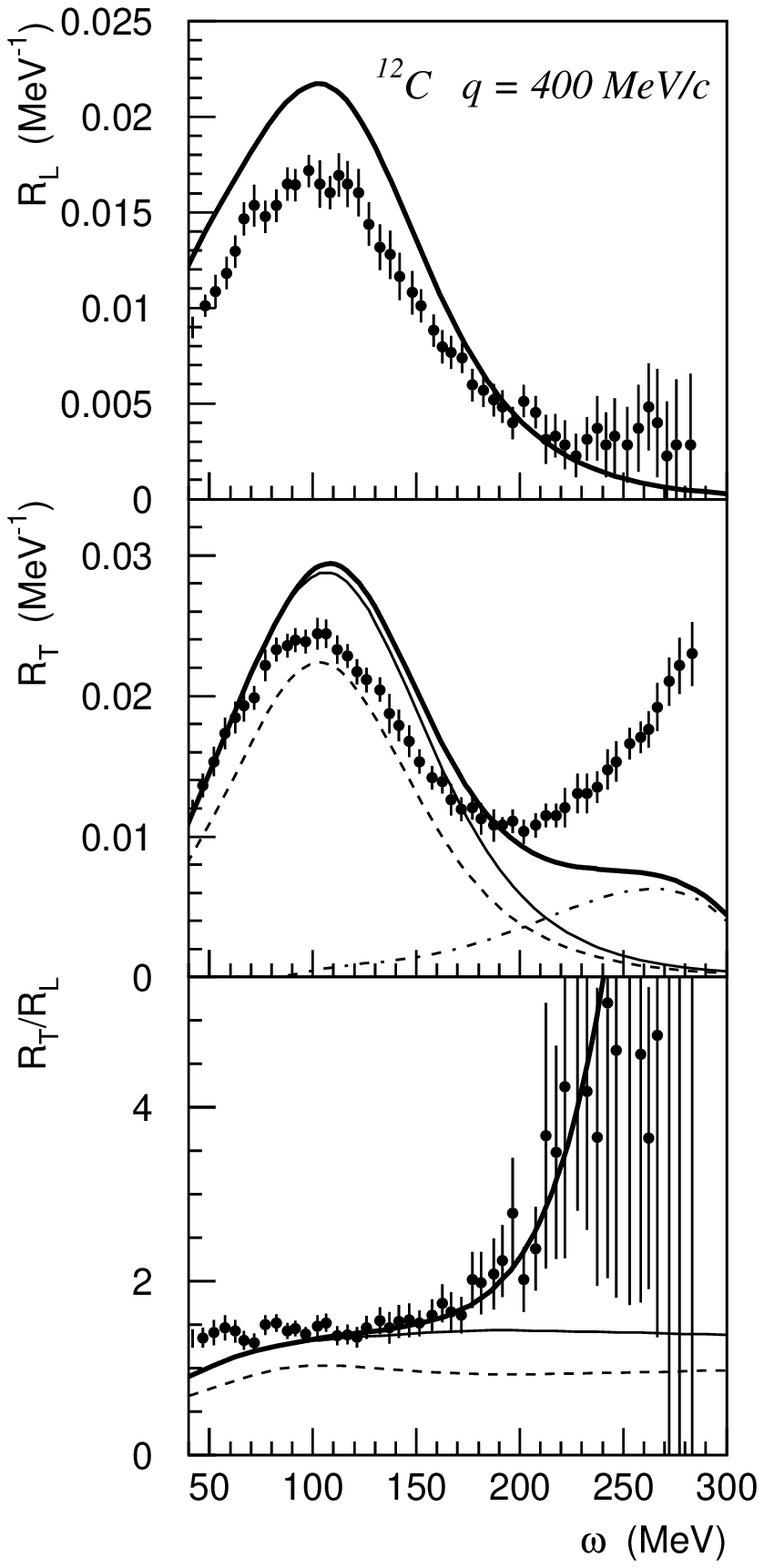}
\caption[3]{The separated structure functions and transverse to
longitudinal ratio ($R_T/R_L$) for ${}^{12}$C(e,e$'$) at $q=400$ MeV$/$c. The
solid line and dashed line is the calculated
one-nucleon knockout contribution with and without inclusion of the
two-body currents. The dash-dotted curve corresponds to two-nucleon
knockout and the thick solid line to the total cross section. The data
are taken from ref. \cite{bar} (Saclay).}
\end{figure}

\begin{figure}
\hspace{3cm}
\epsfxsize=10cm
\epsffile{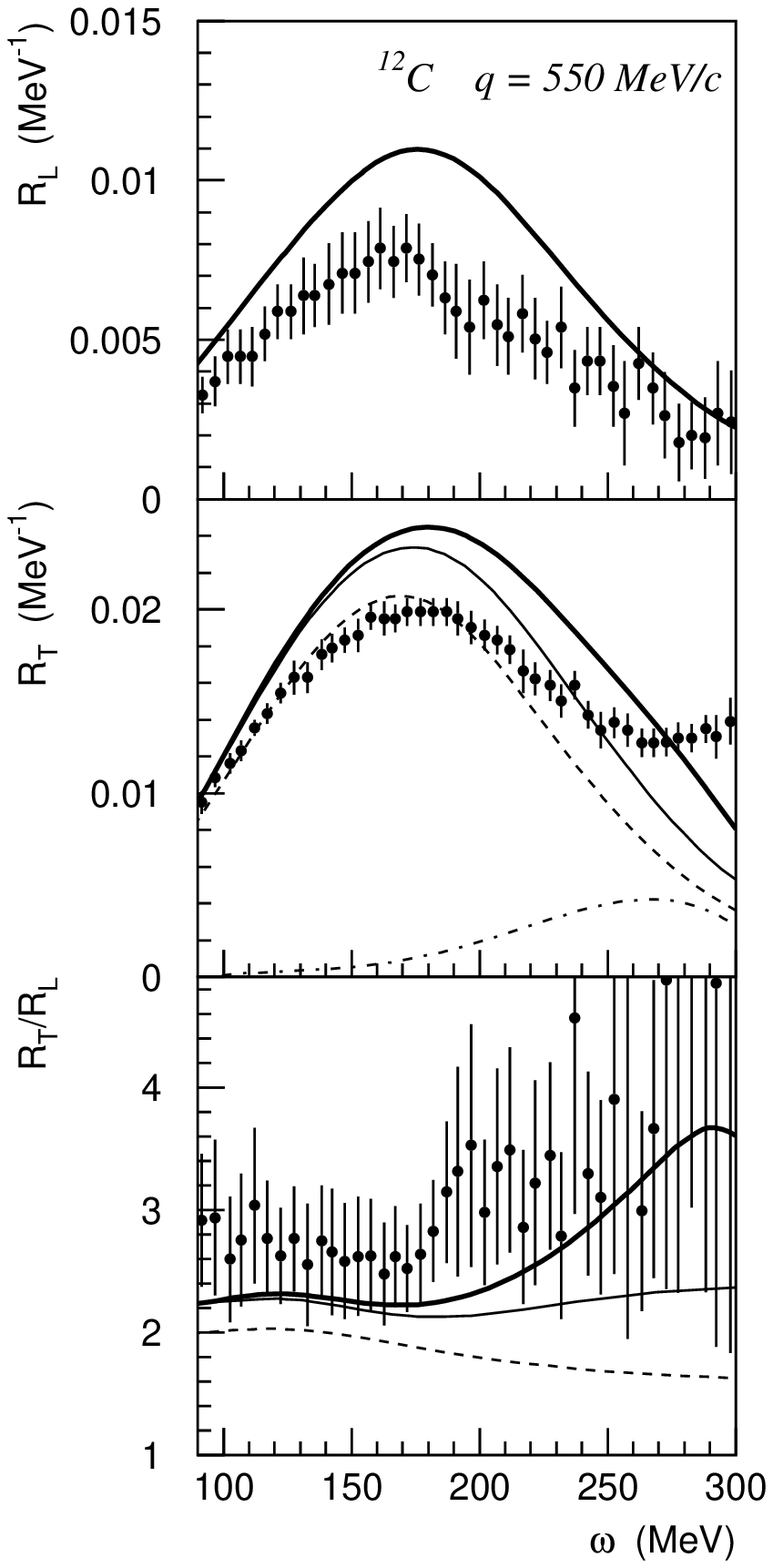}
\caption[4]{ Same as Fig.~2 but for ${}^{12}$C(e,e$'$) at
$q=550$ MeV$/$c.
}
\end{figure}
\begin{figure}
\hspace{3cm}
\epsfxsize=10cm
\epsffile{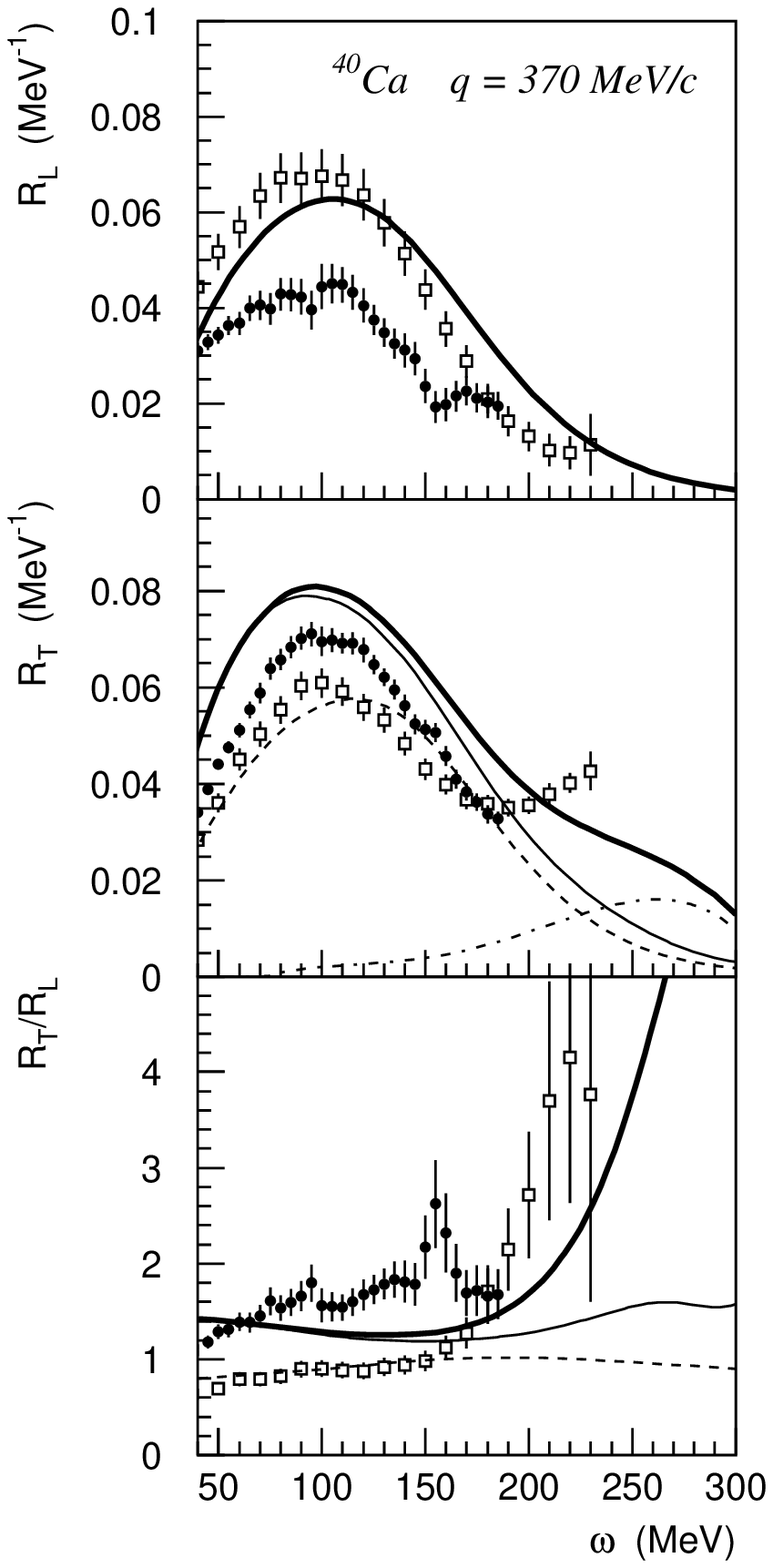}
\caption[5]{ Same line conventions as Fig.~2 but for ${}^{40}$Ca(e,e$'$) at
$q=370$ MeV$/$c.
The data are taken from ref. \cite{mez} (Saclay)
 (dots) and ref. \cite{yat} (Bates)(squares).
}
\end{figure}

\section{Conclusions}

The longitudinal and transverse ${}^{12}$C(e,e$'$) and  ${}^{40}$Ca(e,e$'$)
response
functions have been evaluated in a non-relativistic HF+RPA model including
one- and two-body nuclear currents.
The main goal of this paper was to estimate the
impact of two-body currents on
the inclusive electron scattering structure functions.
The calculations suggest that even in the QE region the two-body currents can
induce an extra $20-30\%$ of strength into the transverse channel.
In the dip region, two-nucleon knockout is found to
gain in relative importance and the two-body currents are predicted to
fill in a large fraction of the missing strength between the IA results and
the data. Therefore, we conclude that, given the degree of importance, two-body
currents play an essential role in any model that aims at a complete
description of inclusive electron scattering from complex nuclei.

\vspace{2cm}

{\bf Acknowledgment}
The authors are grateful to Prof. K.Heyde for fruitful discussions and
suggestions.
We would like to thank Prof. C.F. Williamson and Prof. J. Morgenstern for
kindly providing us with the data files of the $^{40}$Ca(e,e$'$) measurements.
This work has been supported by the National Fund for
Scientific Research (NFWO) and in part by the NATO through the research
grant NATO-CRG920171.

\end{document}